\documentclass[aps,pre,twocolumn]{revtex4-1}
\usepackage{mathtools}%
\usepackage{amssymb}%

\usepackage[utf8]{inputenc}%
\usepackage[T1]{fontenc}%
\usepackage[english]{babel}%

\usepackage{graphicx}
\usepackage{dcolumn}
\usepackage{bm}

\begin{document}
\title{An Ising Model for Metal-Organic Frameworks}
\author{Nicolas H\"oft}
\affiliation{Institut f\"ur Theoretische Physik II, Heinrich
Heine-Universit\"at D\"usseldorf, Universit\"atsstra\ss e 1, 
40225 D\"usseldorf, Germany}
\author{J\"urgen Horbach}
\email{horbach@thphy.uni-duesseldorf.de}
\affiliation{Institut f\"ur Theoretische Physik II, Heinrich 
Heine-Universit\"at D\"usseldorf, Universit\"atsstra\ss e 1, 
40225 D\"usseldorf, Germany}
\author{Victor Mart\'in-Mayor}
\affiliation{Departamento de F\'isica  Te\'orica I, Universidad
Complutense, 28040 Madrid, Spain}
\affiliation{Instituto de Biocomputaci\'on y F\'{\i}sica de Sistemas Complejos (BIFI), 50018 Zaragoza, Spain}
\author{Beatriz Seoane}
\email{beaseobar@gmail.com}
\affiliation{Laboratoire de physique th\'eorique, D\'epartement de physique de
l'ENS, \'Ecole normale sup\'erieure, UPMC
Paris 06, CNRS, PSL Research University, 75005 Paris, France}
\affiliation{Sorbonne Universit\'es, UPMC Univ. Paris 06, \'Ecole normale
sup\'erieure, CNRS, Laboratoire de Physique
Th\'eorique (LPT ENS), 75005 Paris, France}
\affiliation{Instituto de Biocomputaci\'on y F\'{\i}sica de Sistemas Complejos (BIFI), 50018 Zaragoza, Spain}
\begin{abstract}
We present a three-dimensional Ising model where lines of equal spins
are frozen in such that they form an ordered framework structure. The
frame spins impose an external field on the rest of the spins (active
spins). We demonstrate that this ``porous Ising model'' can be seen
as a minimal model for condensation transitions of gas molecules in
metal-organic frameworks. Using Monte Carlo simulation techniques,
we compare the phase behavior of a porous Ising model with that of
a particle-based model for the condensation of methane (CH$_4$) in
the isoreticular metal-organic framework IRMOF-16.  For both models,
we find a line of first-order phase transitions that end in a critical
point. We show that the critical behavior in both cases belongs to the
3D Ising universality class, in contrast to other phase transitions in
confinement such as capillary condensation.
\end{abstract}

\maketitle

\section{Introduction}
\label{sec1}
The Ising model has been a paradigm for the study of phase transitions.
The analytical solution of the two-dimensional (2D) Ising model
allowed for the first prediction of non-mean-field critical exponents
\cite{onsager44}. Monte-Carlo simulations as well as renormalization group
calculations of the three-dimensional (3D) Ising model have provided
very accurate computations of critical exponents that establish the 3D
Ising universality class \cite{ferrenberg91,pelissetto02}. Now, these
exponents are known with very high precision thanks to the conformal
bootstrap \cite{kos16}.  Moreover, for phase transitions in confinement
and in porous media, Ising models have provided a detailed understanding
of wetting phenomena \cite{nakanishi82,ball88,ball89,binder03} as well
as the exploration of novel condensation transitions such as interface
localization-delocalization \cite{binder03,binder08} and the random-field
Ising model universality class in disordered porous media (see, e.g.,
\cite{belanger91} and references therein).

Metal-organic frameworks (MOFs) are a relatively new class of
porous media \cite{li99,eddaoudi02,yaghi03,rowsell05}, in which
gases such as carbon dioxide (CO$_2$), water steam or methane
(CH$_4$) can be stored via condensation on the framework structure
\cite{rowsell05,rosi03,yildirim05,siberio07,uzun14}.  MOFs form a
crystalline porous network where metal-oxide centers are connected
with each other by organic linkers. In this work, we consider the
iso-reticular MOF structure IRMOF-16 in which the metal-oxide
centers consist of an ordered arrangement of ZnO tetrahedra
(see below). Gas condensation on various IRMOFs has been recently
studied via Monte Carlo (MC) simulations in the grandcanonical ensemble
\cite{mueller05,walton07,dubbeldam07,liu08,fairen10,toni10,hicks12,desgranges12,hoeft15,braun15}.
These studies have found evidence for lines of first-order transitions
that end in a critical point. In Refs.~\cite{hoeft15,braun15}, it has
been explicitly shown that for each of these condensation transitions
there is coexistence of bulk phases that extend over the unit cells of the
framework. Moreover, H\"oft and Horbach \cite{hoeft15} have demonstrated
for the condensation of CH$_4$ in IRMOF-1 that there are two lines of
first-order condensation transitions, both ending in a critical point.
The first line at lower densities is associated with a novel type of
phase transition on the surface of IRMOF-1 and has thus been denoted as
IRMOF surface (IS) transition in Ref.~\cite{hoeft15}. The second one, the
IRMOF liquid-gas (ILG) line, can be seen as the analog of the liquid-gas
line in bulk fluids.

Also the thermodynamic properties around the two critical points of
CH$_4$ in IRMOF-1 were studied in Ref.~\cite{hoeft15}. Evidence was given
that the critical behavior belongs to the 3D Ising universality class.
However, especially for the ILG critical point, the situation is not so
clear. Here, the critical point is at a relatively high CH$_4$ density
and thus the acceptance rates for insertion moves in the grand-canonical
Monte Carlo simulation are low. This only allows to consider relatively
small systems for which the corrections to finite-size scaling in terms
of, e.g., the Binder cumulant \cite{binder81} are relatively large.
To circumvent these problems, a minimal model of the Ising type would
be helpful that shows a phase transition similar to the ILG transition
for CH$_4$ in IRMOF-1.  Then, one could more accurately investigate the
critical behavior of the latter transition and rationalize that it is
a member of the 3D Ising universality class.

In this paper, we propose a 3D Ising model where lines of equal spins
are fixed. These lines are arranged such that they form a simple cubic
framework structure. The fixed spins exert a field on the ``mobile''
active spins that tends to align the latter in the direction of the
former.  By compensating this field by a homogeneous external magnetic
field acting on the active spins, coexistence may occur between a phase of
positive and one of negative magnetization.  In fact, via MC simulations
we demonstrate the existence of a line of first order transitions that
end in a critical point and compare the resulting phase diagram to MC
simulations of the condensation of CH$_4$ in IRMOF-16. Different to our
previous study \cite{hoeft15}, we consider IRMOF-16, because it has
larger pores than IRMOF-1 and therefore the coexistence range of the
ILG transition in IRMOF-16 is much broader than the one in IRMOF-1. Note
that we do not find the IS transition line in IRMOF-16. Probably, these
transitions occur at very low temperatures in IRMOF-16 and might be
only metastable since the stable states are expected to be crystalline
in this temperature range.

Both for the porous Ising model and CH$_4$ in IRMOF-16, in the MC
simulations advanced sampling techniques are used, namely tethered and
successive umbrella sampling \cite{fernandez09,virnau04}, respectively.
At a given temperature $T$, these techniques allow to accurately determine
the probability distributions $P({\cal O})$ of a variable ${\cal O}$ that
is directly associated with the order parameter. Note that for the Ising
model and the IRMOF-16 system the variables ${\cal O}$ are given by the
magnetization, ${\cal O}=M$, and the density of the adsorbed gas, ${\cal
O}=\rho$, respectively. Up to a constant $C$, the probability distribution
$P({\cal O})$ corresponds to minus the logarithm of the free energy,
$\beta F({\cal O}) \propto - {\rm ln} P({\cal O}) + C$ (with $\beta$
the inverse thermal energy), and therefore the full information about
the thermodynamics of the system can be obtained from this quantity, in
particular the phase diagram and quantities required for the finite-size
scaling analysis around the critical point such as the Binder cumulant,
the order parameter or the interfacial free energy.  Our results indicate
similar behavior of $P({\cal O})$ for the two considered systems. Compared
to the corresponding bulk systems, in both cases the critical point shifts
to lower temperature and a higher value of ${\cal O}$.  While strong
corrections to 3D-Ising behavior are seen for the IRMOF-16 system, as
expected for the considered small system sizes in this case, for the
porous Ising model our finite-size scaling analysis clearly indicates
a critical behavior according to the 3D-Ising universality class.

The outline of the paper is as follows: In Sec.~\ref{sec2}, we
introduce the models and the details of the simulations. Then, we
present the results in Sec.~\ref{sec3} and finally draw conclusions
in Sec.~\ref{sec4}.

\section{Models and details of simulations and finite-size scaling
analysis}
\label{sec2}
\subsection{Porous Ising model}
\label{sec:ising_model_outline}
In this section, the details of the porous Ising model are described.  The
layout of this section is as follows. In Sec.~\ref{sec:our_ising_model},
we introduce our Ising model. Given the spin/particle analogy that we aim
to establish, we shall be mostly interested in the low temperature phases.
These phases correspond to condensed phases in the particle system. The
use of advanced sampling techniques is required at low temperatures that
we introduce in Sec.~\ref{sec:ising_simulations}.

\begin{figure}
\includegraphics[width=0.35\textwidth]{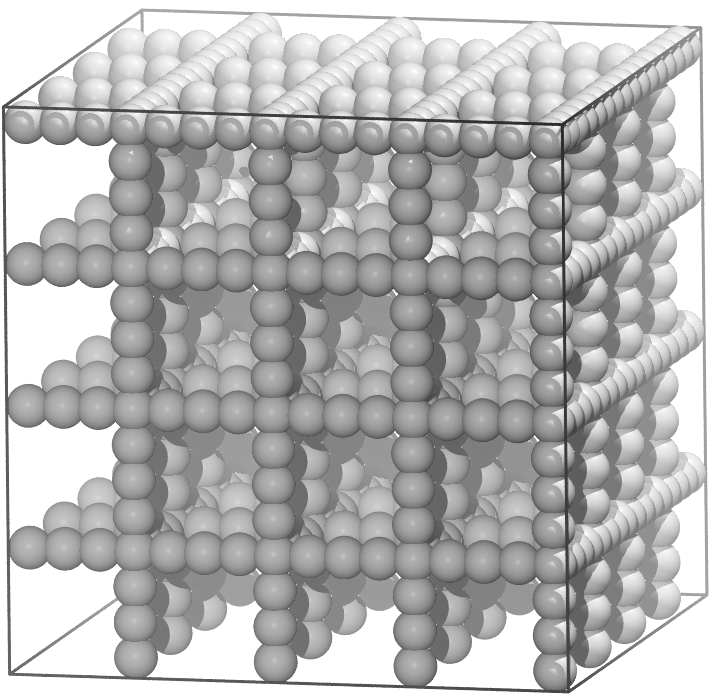}
\includegraphics[width=0.35\textwidth]{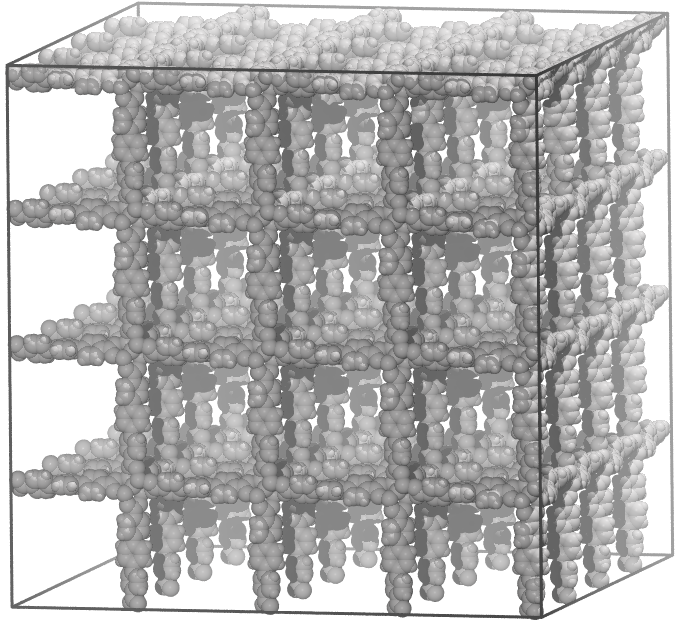}
\caption{\label{fig1} Porous networks of the Ising system showing the
fixed spins only (upper panel) and IRMOF-16 (lower panel). A pore of
the Ising model consists of 10 fixed spins, the IRMOF-16 framework pore
contains 113 atoms.}
\end{figure}
\subsubsection{Model}
\label{sec:our_ising_model}
We consider $N$ Ising spins ($\sigma_i=\pm 1$) on a cubic lattice of
size $L$, endowed with periodic boundary conditions. In this lattice,
we shall distinguish two types of spins: the \emph{active} ones and the
\emph{frame} spins.

The frame spins, depicted in the upper panel of Fig.~\ref{fig1}, mimic the
MOF, see the lower panel in Fig.~\ref{fig1}. The frame is characterized
by a period $P$ (we are assuming that $P$ divides exactly $L$). It is
formed by straight lines parallel to the three lattice axis. We fix all
the spins on the frame to $\sigma_i=1$.

The active spins are our dynamic variables. Their number is
$\mathcal{N}_\mathrm{act}=N(P^3-3P+2)/P^3$, where $N=L^3$. The interaction
energy of the system is given by the exchange term, introducing an
interaction parameter $J>0$, and a coupling to an external magnetic
field $h$
\begin{equation}
\label{eq:hamiltonian}
{\cal H}=-J \sum_{\langle i,j \rangle}\sigma_{i}\sigma_{j}
         -h\sum_{i=1}^N\sigma_{j},
\end{equation}
where $\sum_{\langle i,j \rangle}$ runs over all the couples of nearest
neighbors. We consider all the spins, both frame and active ones,
in Eq.~(\ref{eq:hamiltonian}). Therefore, at variance to the usual 3D
Ising model (that we will refer as the {\em bulk}), the frame induces
an effective positive magnetic field over the active spins even if $h=0$.

The order parameter of this system is linked to the magnetization of
the active spins,
\begin{equation}
{\cal M}=\sum_{i\in \mathcal{N}_\mathrm{act}}\sigma_{i}\,.
\end{equation}
Note that one can interpret the Ising model as a lattice gas: $s_i=1$
(-1) meaning that a particle is present (absent) at site $i$. Therefore,
the magnetization density relates to the particle-number density
straightforwardly:
\begin{equation}
m={\cal M}/\mathcal{N}_\mathrm{act}=2 \rho -1.
\end{equation}
The presence of the fixed sublattice of spins displaces the phase
diagram in the parameter space, as it is shown in Fig.~\ref{fig6}
for the $P=4$ and $J=1$ model. However, the qualitative behavior of
the phase diagram remains unaltered: a first order line that ends in a
critical point separating a paramagnetic and ferromagnetic phases. The
nature of this critical point could, in principle, change because of
the symmetry break imposed by the sublattice. We will show that this
point remains, being universal and in the 3D universality class.

In this work, we have considered three variants of this model: $J=1$
and $P=4$ (Model 1), $J=1$ and $P=8$ (Model 2), and $J=0.1$ and $P=4$
(Model 3).  All the figures shown in this manuscript corresponds
to Model 1. Results for the Models 2 and 3 are summarized in
Table~\ref{tab:criticalpointIsing}.

\subsubsection{Simulation details for the porous Ising model}
\label{sec:ising_simulations}
As we explained above, the magnetic/particle analogy made us focus
on the low temperature phase. In that region, the system undergoes
a first-order phase transition upon varying the applied field,
$h$. Now, the simulation of first-order transitions is intrinsically
difficult~\cite{martin-mayor07}. This is why we shall refer to
a special simulation method, named tethered Monte Carlo. Our
description will be brief (the interested reader may consult
Refs.~\cite{fernandez09,martin-mayor11,fernandez12}).
 
Tethered Monte Carlo is a sophistication of the traditional umbrella
sampling \cite{torrie77}, where the constrained free-energy $W_\beta(\hat
m)$ is reconstructed by means of a (numerically exact) thermodynamic
integration. The constrained free-energy is defined from
\begin{equation}
\mathrm{e}^{- N W_\beta(\hat m)} =
\sum_{\{\sigma_i\}} \mathrm{e}^{-\beta {\cal H} 
- \frac{1}{2} N (\hat m -m)^2} \,. 
\end{equation}
In the above expression, the sum runs over all possible configurations
of the active spins, while $\beta$ stands for the inverse temperature
$1/T$. It is clear from the definition that $W_\beta(\hat m)$ is the
constrained free energy needed to keep the system at a magnetization $m
(={\cal M}/N_\mathrm{act})\approx \hat m$ at temperature $T$ and at zero
external magnetic field.

From $W_\beta(\hat m)$, one can trivially recover the canonical partition
function in a magnetic field as
\begin{equation}
\int {\rm d} \hat m \; \mathrm{e}^{- N W_\beta(\hat m)+ N \beta h \hat m} =
Z(h,T) \, \mathrm{e}^{\frac{1}{2} N\beta^2 h^2}\,.
\end{equation}
It follows that the tethered parameter $\hat m$ and the magnetization
density are related as:
\begin{equation}
\langle \hat m \rangle_{\beta,h} = 
\langle m \rangle_{\beta,h} + \beta h \,.
\end{equation}
What one actually computes in a tethered computation is the derivative
with respect to $\hat m$
\begin{equation}
W_\beta'(\hat m) = \langle \hat m -m\rangle_{\hat m,\beta}\,.
\end{equation}
Therefore, we numerically compute $W_\beta'(\hat m_i)$ on a grid
$(\hat m_i,T)$, with $i=1,\ldots N_m$. The entire potential can later be
recovered by means of a numerical integration of these points. From this,
we determine very precisely the location of the first-order transition at
a given temperature $T$, that is, the coexistence field $h_\mathrm{co}$
and the position of the positive $\hat m^+$ and negative $\hat m^-$
magnetization minima of the total free-energy potential. This amounts
to performing a Maxwell construction~\cite{martin-mayor07}:
\begin{equation}\label{eq-hco}
\log \frac{p_{\beta,h}(\hat m^+)}{p_{\beta,h}(\hat m^-)} =
-N\int_{\hat m_-}^{\hat m^+} 
\left(W_\beta'(\hat m)-h_\mathrm{co}\right) {\rm d} \hat m = 0.
\end{equation}
$W_\beta'(\hat m)-h_\mathrm{co}$ has at least three roots: the
magnetization of the two pure phases $\hat m^-$, $\hat m^+$, and a
central point magnetization, $\hat m^*$, that corresponds to a half-half
configuration of the two ferromagnetic phases. We can use this fact to
obtain the free-energy cost to build the two interfaces (because of the
periodic boundary conditions) of size $L^2$, by comparing the free-energy
of the mixed configuration with $\hat m=\hat m^*$, and of the pure
phase. Once the cost in free-energy is known, the surface tension follows:
\begin{equation}\label{eq:Sigma-def}
\Sigma_\beta=\frac{N}{2L^2}
\int_{\hat m^-}^{\hat m^*}
\left(W_\beta'(\hat m)-h_\mathrm{co}\right) {\rm d} \hat m .
\end{equation}
Alternatively, one can also obtain the first-order transition line,
as well as to expand it into the paramagnetic phase (which is known
as the Widom line~\cite{widom65}), by looking for the value of $h$
that makes $p_\beta(\hat m;h)$ symmetrical, in practice, by extracting
the value of $h$ at which the skewness of the probability distribution
vanishes~\cite{parisi14}. This approach allows us to compute the Binder
cumulant \cite{binder81}, $U_L$, for a given linear dimension $L$ of 
the simulation box as the kurtosis of the distribution along this line,
\begin{equation}
\label{eq_binder_ising}
U_L(T,h) = 1-\frac{1}{3}
\frac{\langle\,( \hat m- \langle \hat m \rangle_{\beta,h} )^4\,\rangle}
{\langle\,( \hat m- \langle \hat m \rangle_{\beta,h} )^2\,\rangle^2} \, .
\end{equation}

Before we go on a word of caution is in order. Dimensionless quantities
such as $U_L$ or the surface tension $L^2\Sigma_\beta$, are expected to
enjoy only a \emph{restricted} degree of universality at the critical
point.  Their value is expected to be independent of any microscopic
details of the interactions, however they are sensitive to several
geometric features. For instance, changing boundary conditions or
the lattice geometry (say, going from a cubic to an elongated box)
must result in a variation of their value. The question arose in the
original paper by Binder~\cite{binder81}, and has been throughly studied
numerically in bulk systems~\cite{selke06,selke07,selke09}. The reason
for this geometric sensitivity is particularly clear for the Binder
cumulant $U_L$, which can be computed from space integrals of universal
scaling functions~\cite{salas00}. The scaling functions themselves
are insensitive to microscopic details such as the interaction range,
etc. Yet, the value of the integrals that yield $U_L$ are sensitive to
the geometry of the integration domain.

We have performed two sets of simulations, a coarse one to determine
the position of the first-order transition branches and to get a rough
idea of the position of the critical point, and an extensive study of
the critical point. For the first part of the study, we used a mesh
of $m$ points with a width of $\delta m=0.1$, while for the second
part, we reduced this width to $\delta m=0.003$.  For determining the
first-order transition lines, we performed simulations at different
temperatures and $N=16,\ 32 $ and $64$, while for the critical point
we just simulated one temperature for each model $T_\mathrm{sim}\sim
T_\mathrm{c}$, $T_\mathrm{sim}$ is shown in Table~\ref{tab:ana}, and
extrapolated results to nearby temperatures using the re-weighting
method~\cite{ferrenberg88}. In order to compute the critical point
and its critical exponents shown in Table~\ref{tab:ana}, we used the
quotients method~\cite{amit05,ballesteros00}, for which we studied the
crosses of the curves of $\Sigma L ^2$ and $\Delta m=\hat m^+-\hat m^-$
between curves coming from systems at system sizes $L$ and $2L$. With
this scheme, we studied $L=8,\ 12,\ 16,\ 24,\ 32,\ 48$ and $64$.

A difficulty we encounter in the present setting is that the phase
diagram is two-dimensional $(T,h)$. We shall eliminate one variable by
fixing the magnetic field to its coexistence value $h_\mathrm{co}(T)$,
see Eq.~(\ref{eq-hco}).

\subsection{CH$_4$ in IRMOF-16: Model and simulation details}
\begin{figure}
\includegraphics[width=0.3\textwidth]{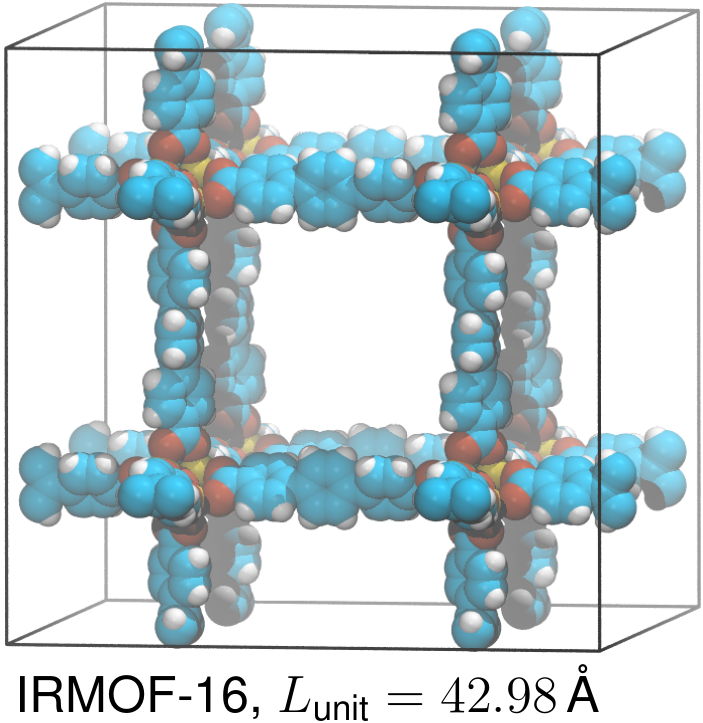}
\caption{Unit cell of IRMOF-16. The length $L_{\rm unit} =
42.980$\,\AA~corresponds to the linear dimension of the unit cell.
Atoms are shown as spheres with different colors, namely C (turquoise),
H (white), O (red), and Zn (yellow). \label{fig2}}
\end{figure}

IRMOF-16 is modeled as a rigid framework, consisting of carbon (C),
hydrogen (H), zinc (Zn), and oxygen (O) atoms. The information about
the relative positions of these atoms are taken from X-ray diffraction
data \cite{li99} (see Figs.~\ref{fig1}b and \ref{fig2}). CH$_4$
molecules are described as Lennard-Jones (LJ) point particles, as
proposed by Martin and Siepmann \cite{martin98}. Also the interactions
of the CH$_4$ particles with the framework atoms are modeled by LJ
potentials, employing the universal force field (UFF) of Rapp\'e {\it
et al.}~\cite{rappe92}. Details on the interactions parameters can be
found in Ref.~\cite{hoeft15}.

The MC simulations for the particle-based model of CH$_4$ in IRMOF-16 are
performed in the grand-canonical ensemble, i.e.~at constant volume $V$,
temperature $T$, and chemical potential $\mu$. The chemical potential
is the analog of the external magnetic field $h$ in the porous Ising
model.  While the field $h$ is the thermodynamic conjugate variable
of the total magnetization ${\cal M}$, the chemical potential $\mu$ is
thermodynamically conjugate to the number of CH$_4$ particles $N$.  Thus,
by changing the intensive variables $h$ in case of the Ising model and
$\mu$ in case of the CH$_4$ in IRMOF-16, the average magnetization ${\cal
M}$ and the average particle number $N$, respectively, can be varied; in
particular, at a given temperature below the critical temperature $T_c$,
the intensive variables $h$ and $\mu$ can be tuned such that coexistence
conditions are obtained.  To this end, histogram reweighting is also used
for the particle-based model, as describe above for the Ising model.
Similarly to the tethered MC, used for the Ising model, the MC for the
IRMOF system is combined with successive umbrella sampling \cite{virnau04}
which allows for an accurate estimate of the probability distribution
$P(\rho)$ in the two-phase region (with $\rho$ the number density of
CH$_4$ particles, $\rho = N/V$). Details on the implementation of the
grand-canonical MC in combination with successive umbrella sampling for
MOFs are given in a previous publication \cite{hoeft15}.

Simulations for different system sizes in a cubic box geometry are
performed. The considered linear dimensions of the boxes are $L_{\rm
MOF} = 1.0$\,$L_{\rm unit}$, $1.5$\,$L_{\rm unit}$, and $2.0$\,$L_{\rm
unit}$, with $L_{\rm unit}= 42.980$\,\AA~the size of the unit cell (see
Fig.~\ref{fig2}). Low acceptance probabilities of the order of $10^{-3}$
for trial insertions of CH$_4$ particles did not allow the simulation
of larger system sizes.  To improve the statistics, for all the systems
10 independent runs were done at each temperature.

\subsection{Results}
\label{sec3}
The central quantity, obtained from the MC simulations for the two
models, is the order parameter distribution function $P({\cal O})$.
Under coexistence conditions of a first-order phase transition,
this function becomes bimodal such that two peaks, located at ${\cal
O}^{(1)}$ and ${\cal O}^{(2)}$ and with equal area under both peaks,
occur \cite{binder84,borgs90}. To obtain the coexistence field $h$ in
case of the porous Ising model and the coexistence chemical potential
in case of the IRMOF system, we employ histogram reweighting techniques
\cite{ferrenberg88}.

\begin{figure}
\includegraphics[width=0.48\textwidth]{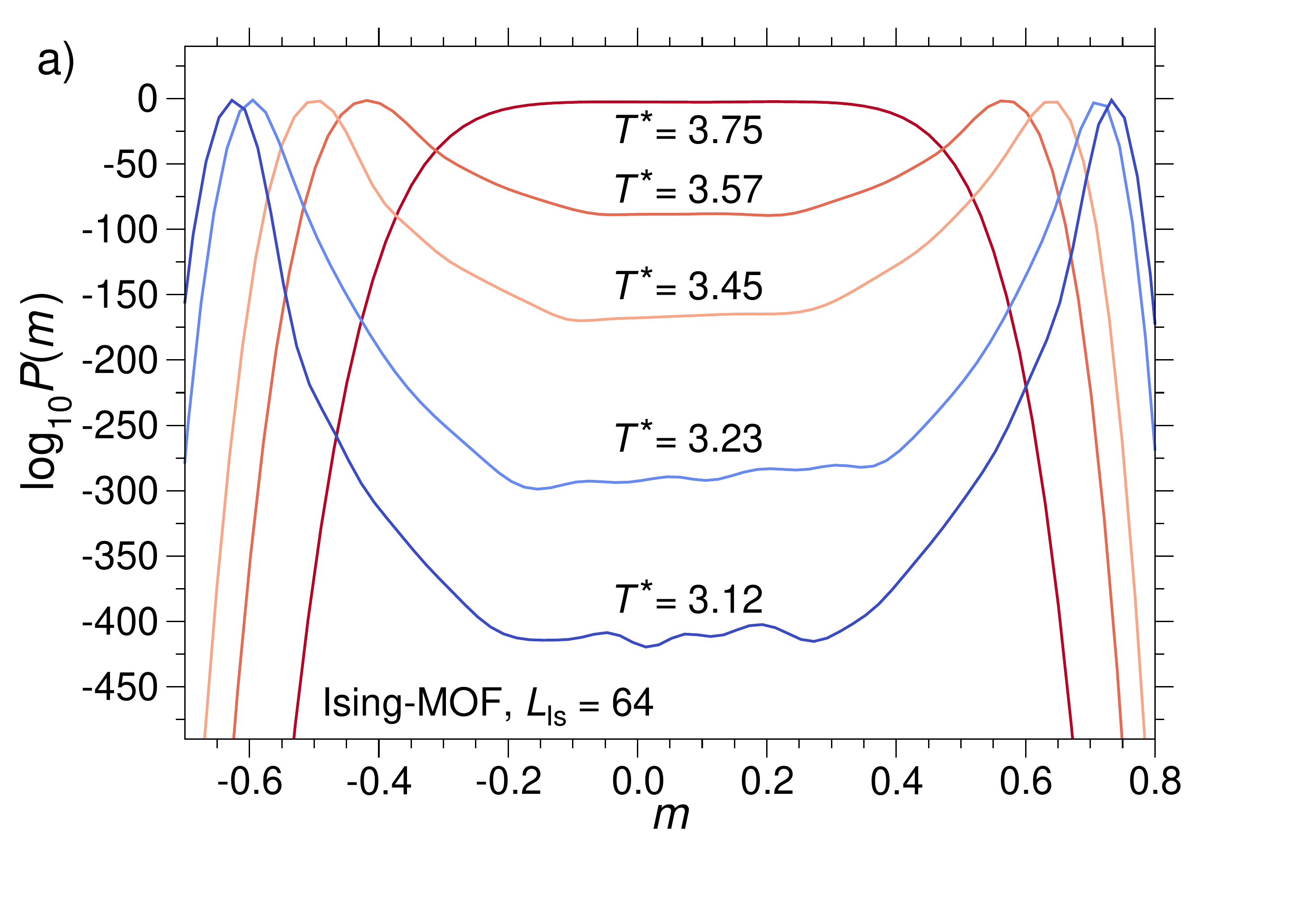}
\includegraphics[width=0.48\textwidth]{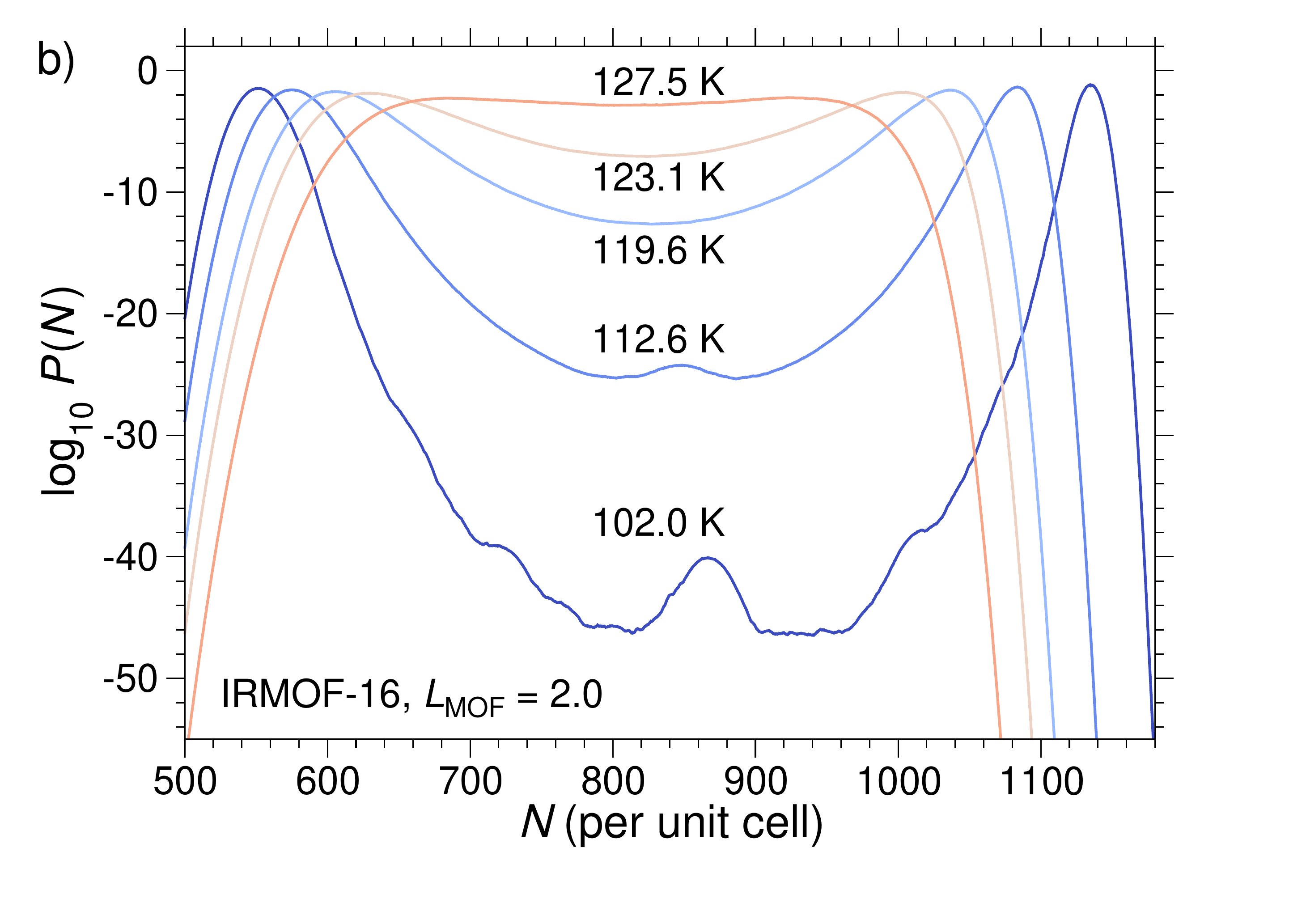}
\caption{\label{fig3} Probability distributions for a) Ising-MOF model and
b) in IRMOF-16 model as determined from the largest available systems,
as indicated by the edge lengths of cubic system sizes, $L_{\rm Is}$
and $L_{\rm MOF}$.}
\end{figure}
Figure \ref{fig3} displays the logarithm of coexistence probability
distributions for the lattice model and the particle-based systems at
different temperatures below the critical temperature.  Note that the
figures show data for the largest systems, simulated in the respective
cases.  For the Ising model, two peaks can be seen at each temperature
and the distance between the peak maxima decreases as approaching the
critical temperature.  Between the peaks, there is a plateau region,
developing ripples for low temperatures, $T \lesssim 3.23 k_BJ$. The
plateau in $P(m)$ corresponds to the two-phase region where the coexisting
phases with magnetization $m^{+}$ and $m^{-}$ are separated from each
other by a planar interface (cf.~the snapshot, Fig.~\ref{fig5}a). The
distance between the height of the peaks and the height of the plateau are
proportional to the surface tension, $\Sigma$, required for the formation
of an interface. The ripples indicate a dependence of $\Sigma$ on $m$ in
the two-phase region. We will clarify the source of this behavior below.

For comparison, the probability distribution $P(N)$ of the ILG transition
of CH$_4$ in IRMOF-16 (Fig.~\ref{fig3}b) shows a similar behavior as the
corresponding function for the Ising model. However, one has to keep
in mind that the considered system size for the atomistic model is much
smaller the one for the Ising model. While the IRMOF-16 system consists of
$2^3$ unit cells or 64 pores, systems with $64^3$ pores are simulated
in case of the Ising model. As we can infer from the distributions in
Fig.~\ref{fig3}b, the oscillations in the regions between the two peaks
are much more pronounced for the IRMOF-16 system and, as we shall see
now, this is due to the much smaller system size considered in this case.

\begin{figure}
\includegraphics[width=0.48\textwidth]{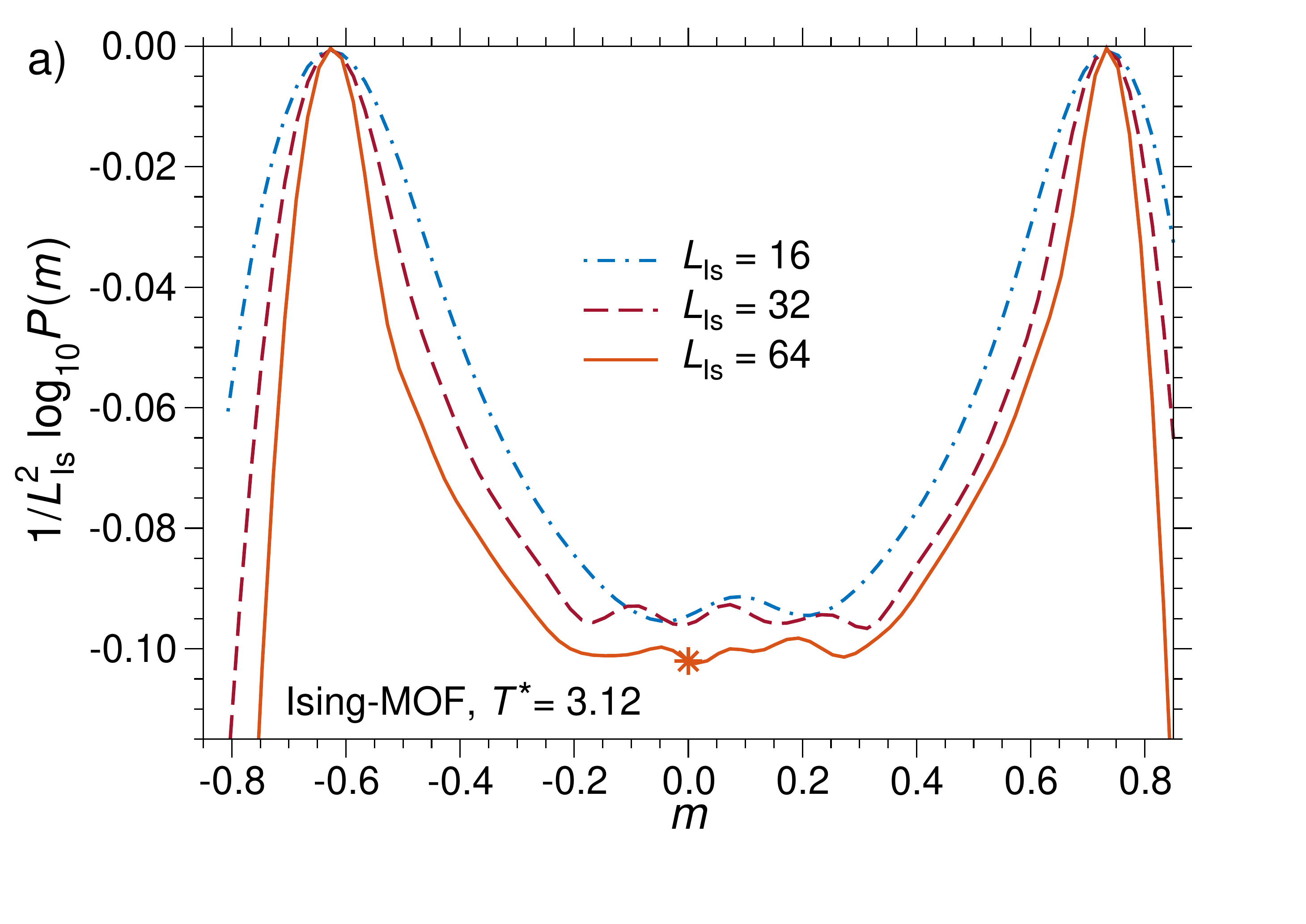}
\includegraphics[width=0.48\textwidth]{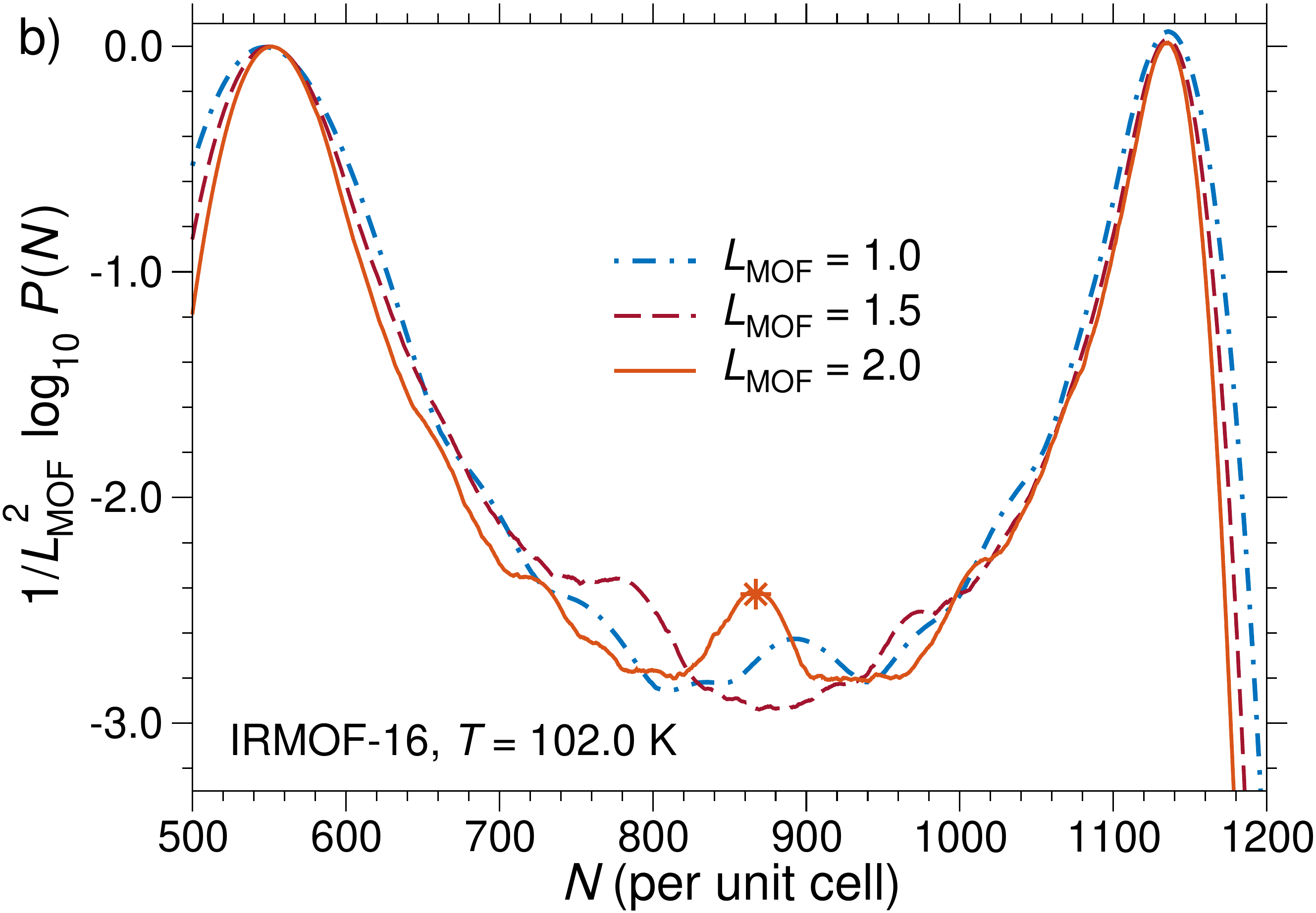}
\caption{\label{fig4} Semilogarithmic plots of the probability
distributions for different system sizes scaled with the area of the
planar interface, $1/L_{\rm Is}^2$ and $1/L_{\rm MOF}^2$ for a) the
Ising model and b) the IRMOF-16 system, respectively. Red stars in a)
and b) refer to the corresponding snapshots in Fig.~\ref{fig5}.}
\end{figure}
\begin{figure}
\centering
\includegraphics[width=0.35\textwidth]{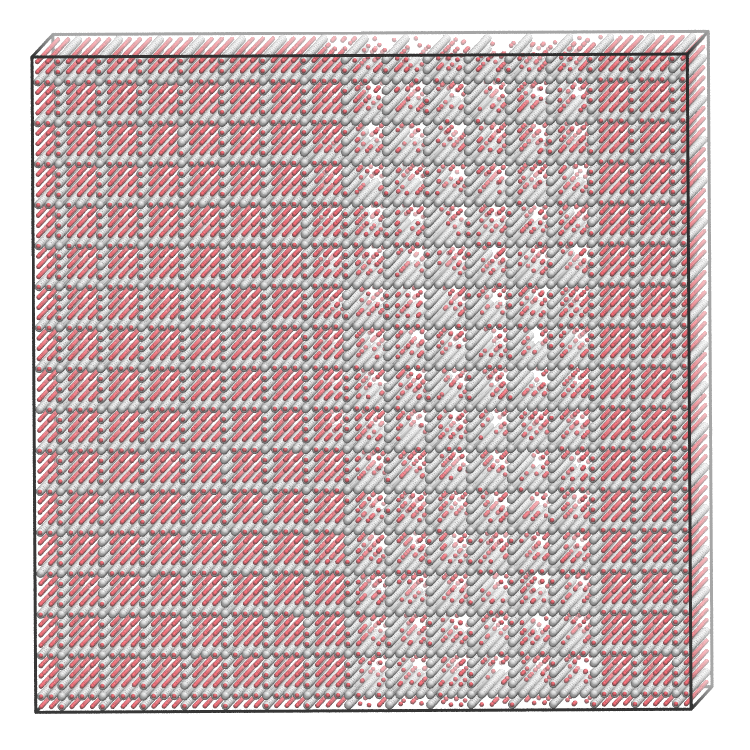}
\includegraphics[width=0.35\textwidth]{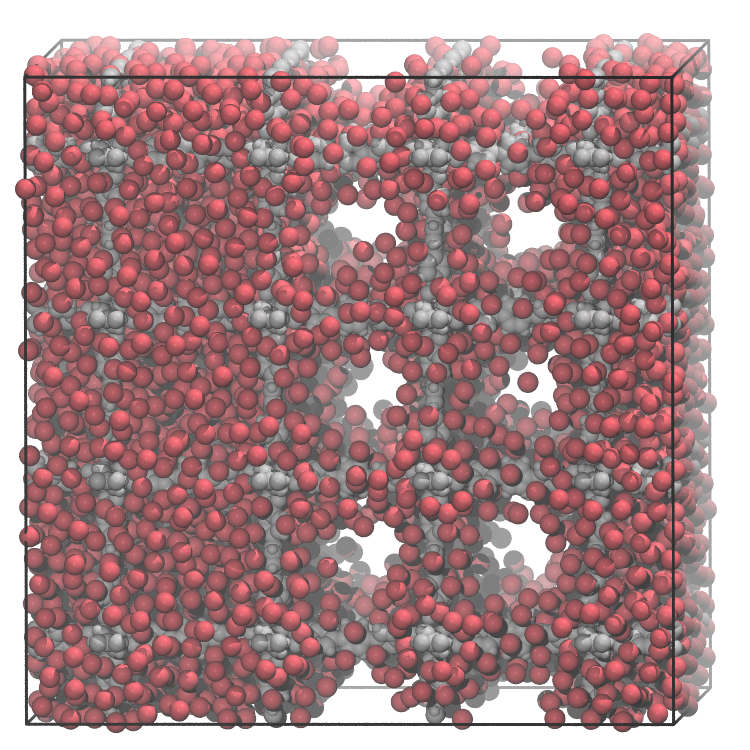}
\caption{\label{fig5} Configuration snapshots at coexistence conditions
for the Ising-MOF model at $T=3.12k_B/J$ (upper panel), no magnetization
($m=0$) showing spins with $\sigma = 1$ only (active spins in red,
fixed framework spins in gray); and the IRMOF-16 model, at $T=102$\,K
with $N=864$ per unit cell (lower panel). Red and gray spheres represent
CH$_4$ molecules and framework atoms, respectively.}
\end{figure}
To this end, we scale the logarithm of the probability distributions by
the area of the interface, $L_{\rm Is}^2$ and $L_{\rm MOF}^2$ for
the Ising model and the atomistic MOF system, respectively. In Fig.~\ref{fig4},
scaled distributions for different system sizes are plotted, for both
the lattice model and the atomistic system at a temperature far below the
critical temperature, corresponding to the lowest temperatures shown
in Fig. \ref{fig3}. As one can infer from $P(m)$ in Fig.~\ref{fig4}a,
with increasing system size, the width of the two peaks decreases while 
the flat region between the two peaks becomes broader. Moreover, the 
distance between maxima and the minimum in $P(m)$ is slightly increasing
with increasing system size. This is due to the fact that in the smaller
system the two interfaces are not sufficiently separated from each other
and thus the interaction between the two interfaces leads to an effective
decrease of the free energy cost of the interfaces. Also the oscillations
in the plateau region of $P(m)$ are less pronounced for the large system.
That these oscillations are expected to vanish in the thermodynamic limit
can be understood as follows: In the two-phase region, the lever rule 
controls the amount of the two coexisting phases with magnetizations
$m^+$ and $m^-$. Under this constraint, in a finite system only for 
certain values of $m$, flat interfaces can be embedded into the framework
structure such that its free energy cost is minimized. This happens when 
the flat interface is located in a plane that goes through the corners of 
the framework structure (cf.~the snapshot for the system with $L_{\rm Is}$, 
Fig.~\ref{fig5}a, corresponding to a minimum in the plateau region of $P(m)$, 
marked by the star in Fig.~\ref{fig4}a). In the thermodynamic limit, a flat
interface can be always arranged according to a minimal free energy cost
and thus the oscillations tend to disappear for sufficiently large 
system sizes. This is also the case if the width of the interfacial region
is of the order of the linear dimension of the unit cell of the framework,
as is expected at sufficiently high temperatures, i.e.~close enough to
the critical temperature. Indeed, as Fig.~\ref{fig4} indicates for the
largest system, this happens for temperatures that are about 10-20\%
below the critical temperature which is around 3.75 (see below).

The scaled distributions for the IRMOF-16 system (Fig.~\ref{fig4}b)
exhibit a similar behavior. However, due to the small system sizes,
finite-size effects are much more pronounced. As one can infer from the
snapshot (Fig.~\ref{fig5}b), even for the IRMOF-16 system with $L_{\rm
MOF}=2.0$ the distance between the two interfaces is less than the linear
dimension of the unit cell. Therefore, $\log P(N)$ shows very pronounced
oscillations in the two-phase region.

\begin{figure}
\includegraphics[width=0.48\textwidth]{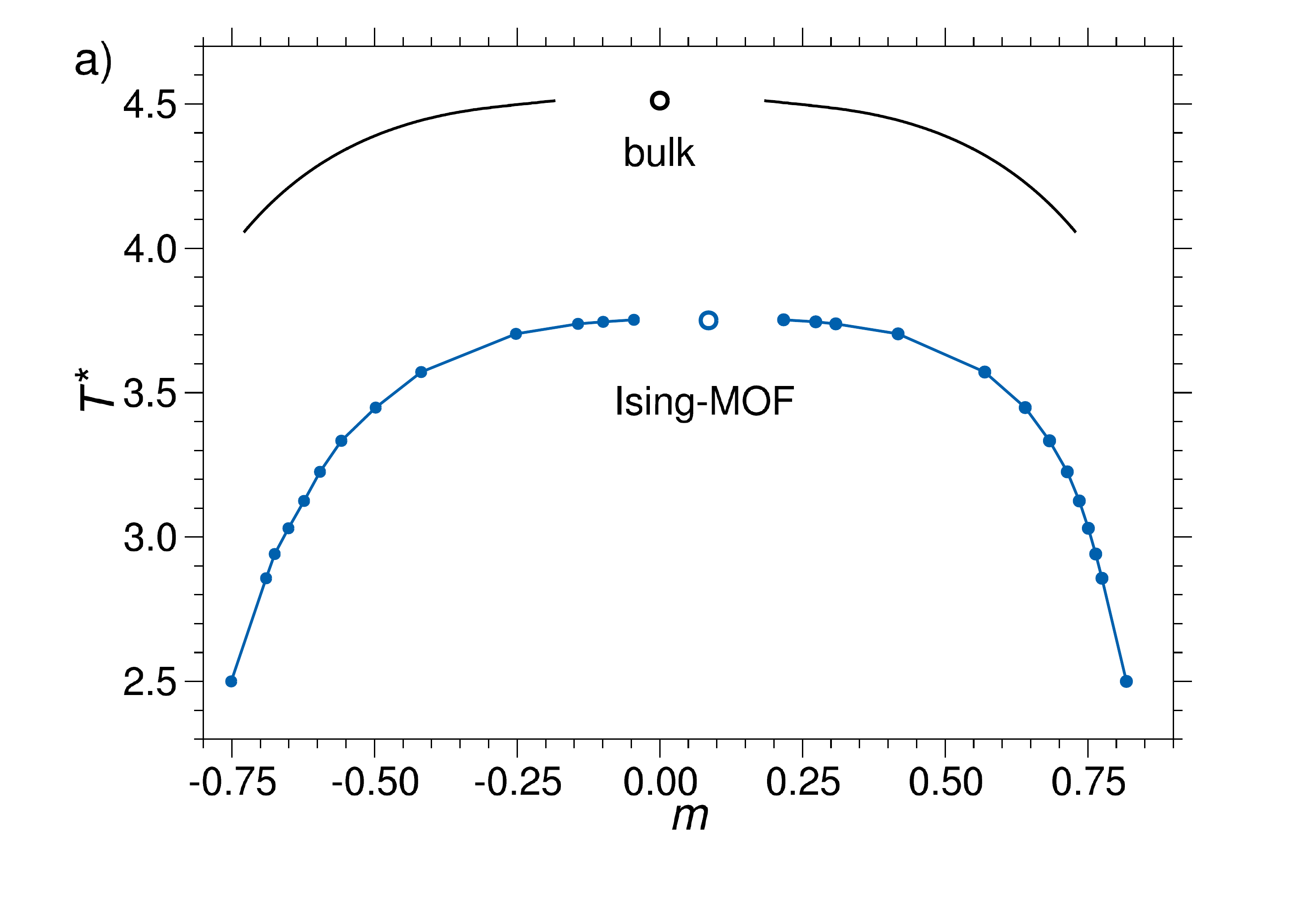}
\includegraphics[width=0.48\textwidth]{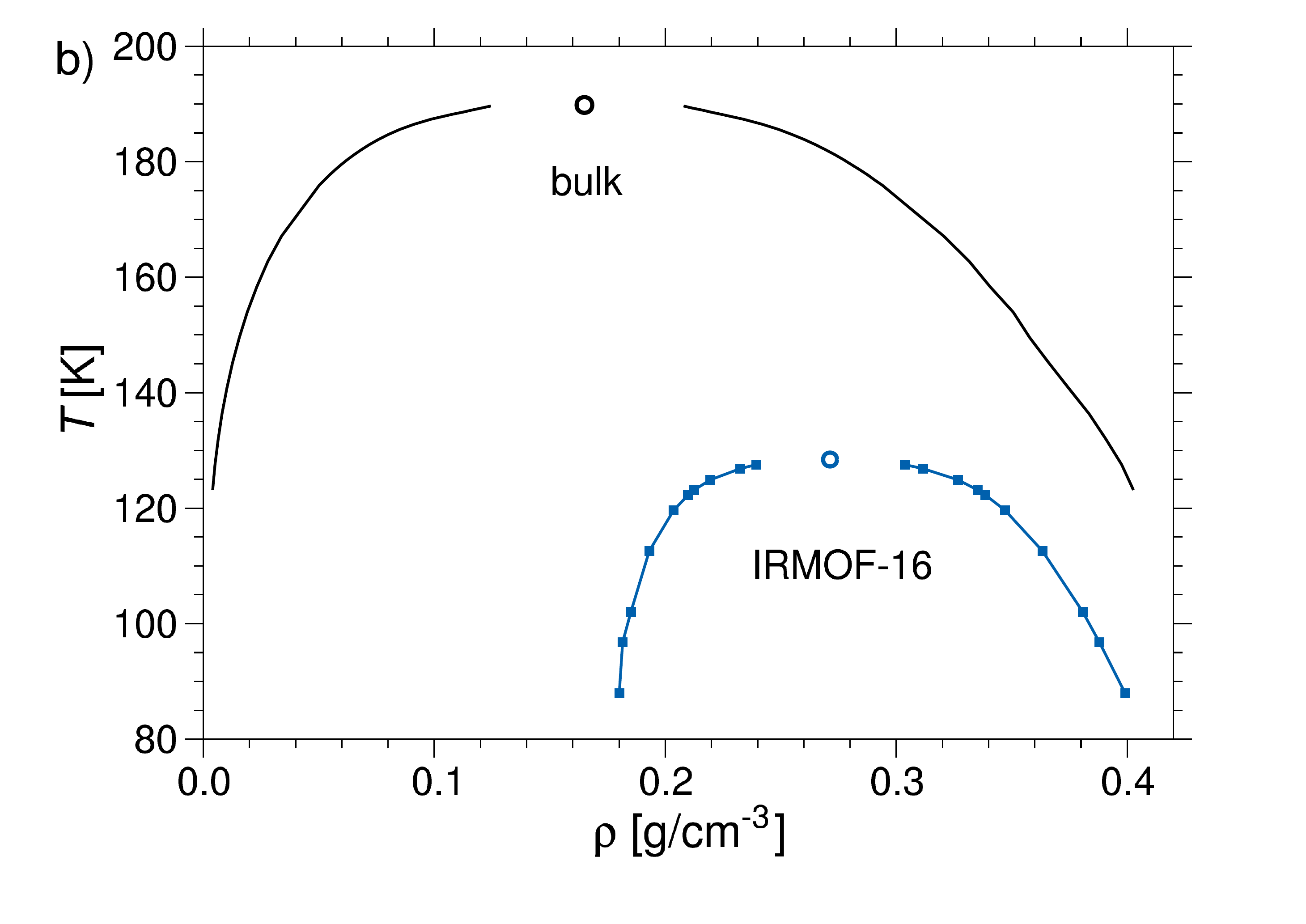}
\caption{\label{fig6} Phase diagram showing binodal lines of a) the Ising
system and b) CH$_4$ in IRMOF-16, showing the ILG phase transition,
each in comparison to the bulk binodal. Critical points are shown as
open symbols.}
\end{figure}
The phase diagrams in the magnetization-temperature and the
density-temperature plane for the porous Ising model and CH$_4$-IRMOF-16,
respectively, are shown in Fig.~\ref{fig6}. Here, the coexistence
magnetizations and densities were directly determined from the first
moments of each of the peaks in $P(m)$ and $P(N)$, respectively. Also
included in the figure are the phase diagrams for the corresponding bulk
systems. Compared to the bulk, in both porous systems the critical
temperature $T_c$ is significantly lower. An analogous effect is
also known from capillary condensation in thin films \cite{fisher81}
and, similarly as in thin films, it is due to the attraction of gas
molecules by the framework structure for the atomistic system and the
alignment of the active spins with the framework spins in case of the
porous Ising model.  Note that the critical temperature of the Ising
system could potentially be tuned to match the behavior of bulk methane
compared to methane in IRMOF-16.  This could be accomplished by varying
the interaction strength of the active spins with the framework spins.

The two-phase region is narrower in the presence of the
framework structure which is also similar to systems in thin
films confinements. However, in thin films, a crossover from 3D to
2D-Ising scaling behavior close to the critical point is observed
\cite{nakanishi82}. Such a cross-over is absent in our case, as we
demonstrate now by a detailed finite-size scaling analysis.

\begin{figure}
\includegraphics[width=0.48\textwidth]{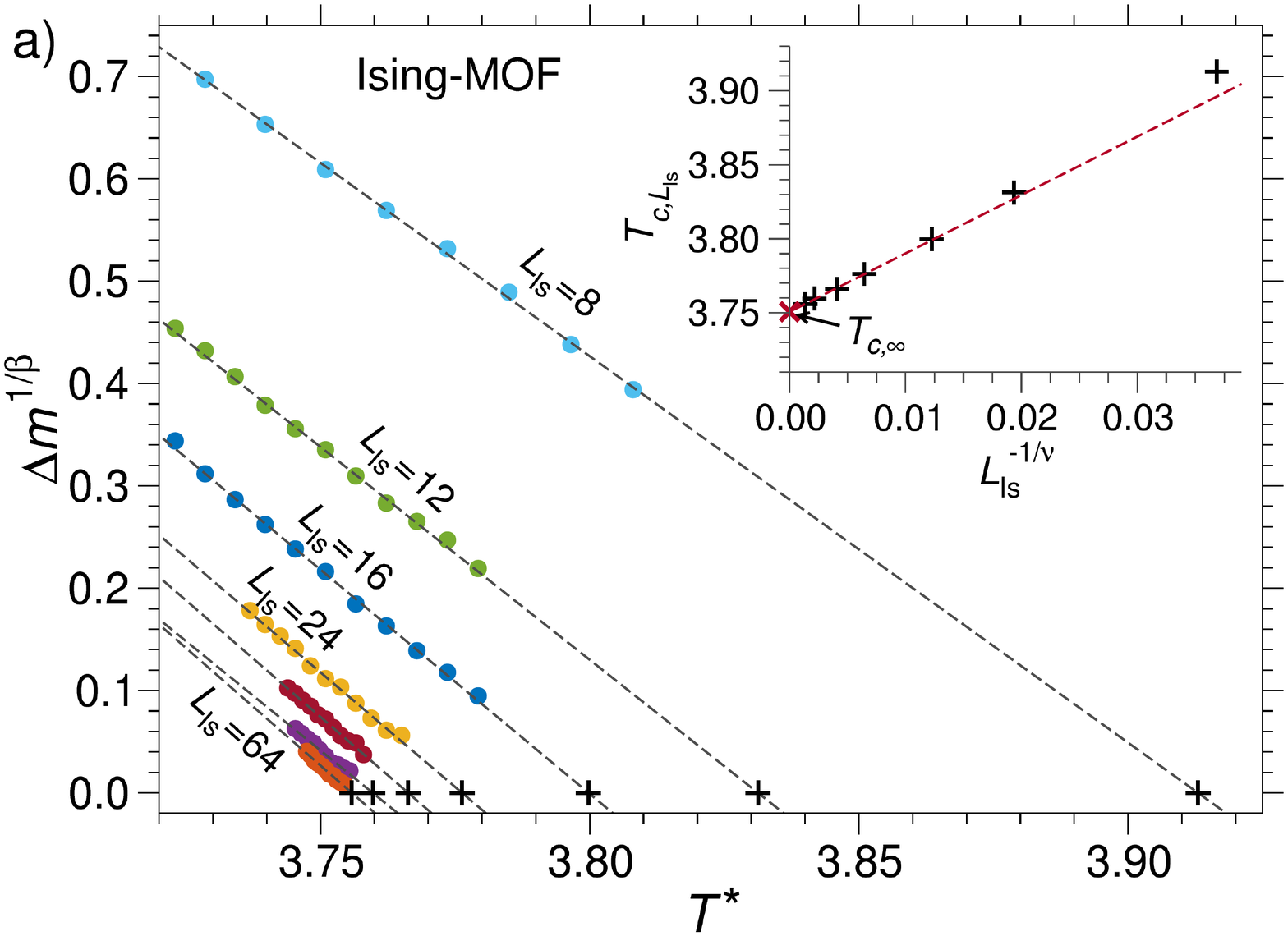}
\includegraphics[width=0.48\textwidth]{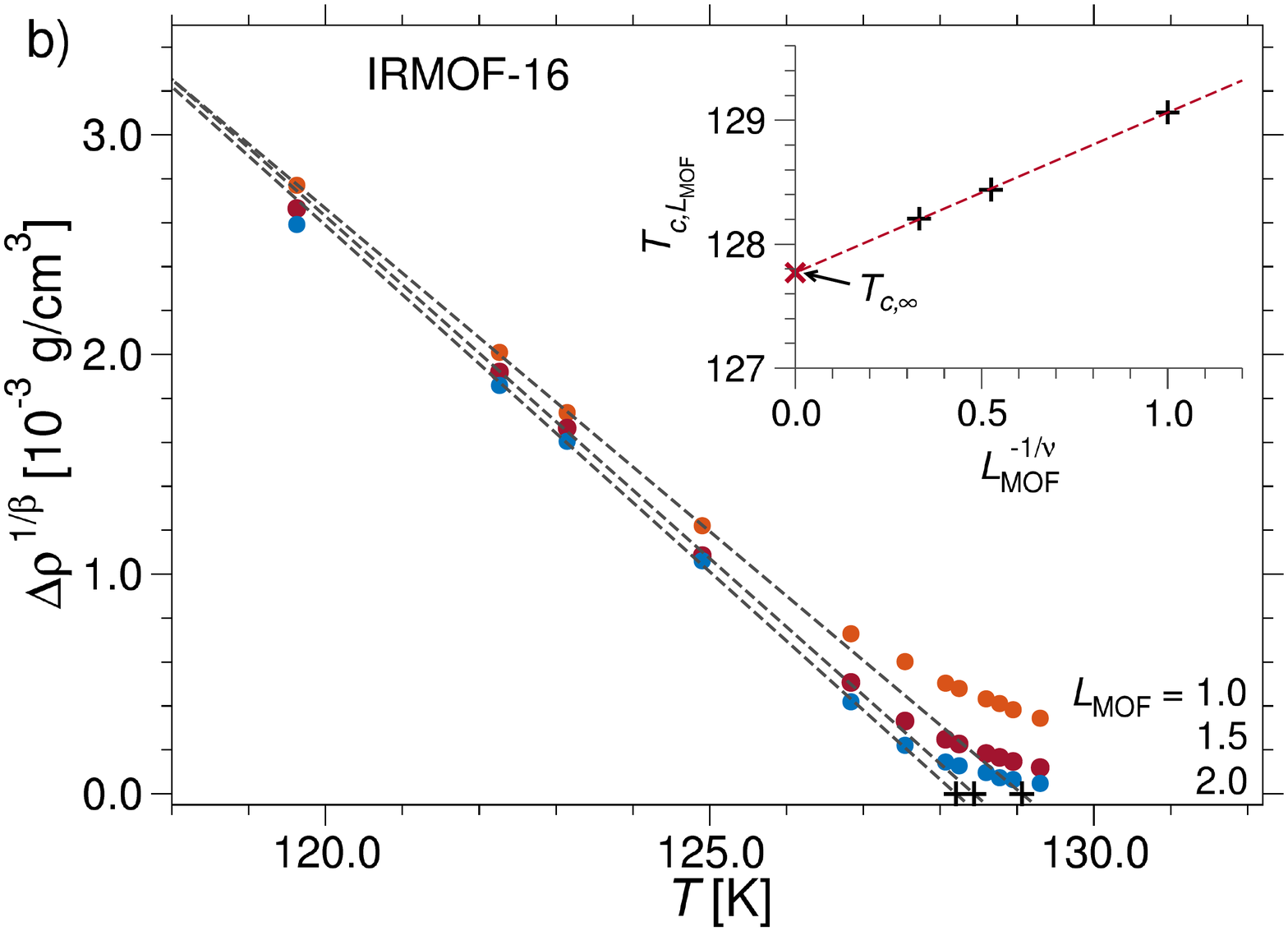}
\caption{\label{fig7} Rectification plot of the order parameter ${\cal
O}$ for a) the Ising model (with ${\cal O}=\Delta m$) and b) the IRMOF-16
system (with ${\cal O}=\Delta \rho$). The insets show $T_c^{L_{\rm Is}}$
and $T_c^{L_{\rm MOF}}$ as a function of $L_{\rm Is}$ and $L_{\rm MOF}$,
respectively. For the exponents, the 3D Ising values $\beta = 0.326$
and $\nu=0.63$ \cite{hasenbusch10} are used.}
\end{figure}
\begin{figure}
\includegraphics[width=0.48\textwidth]{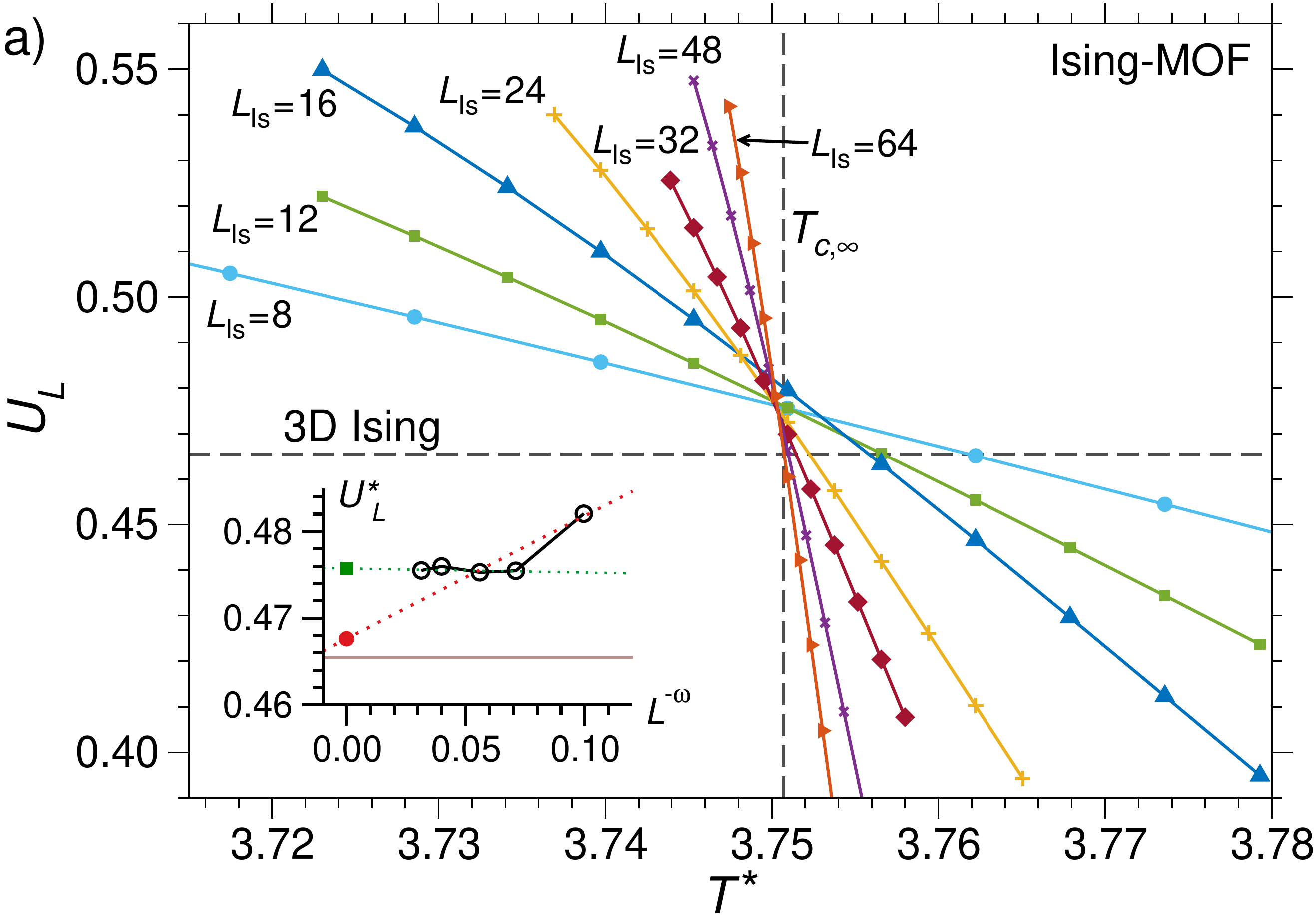}
\includegraphics[width=0.48\textwidth]{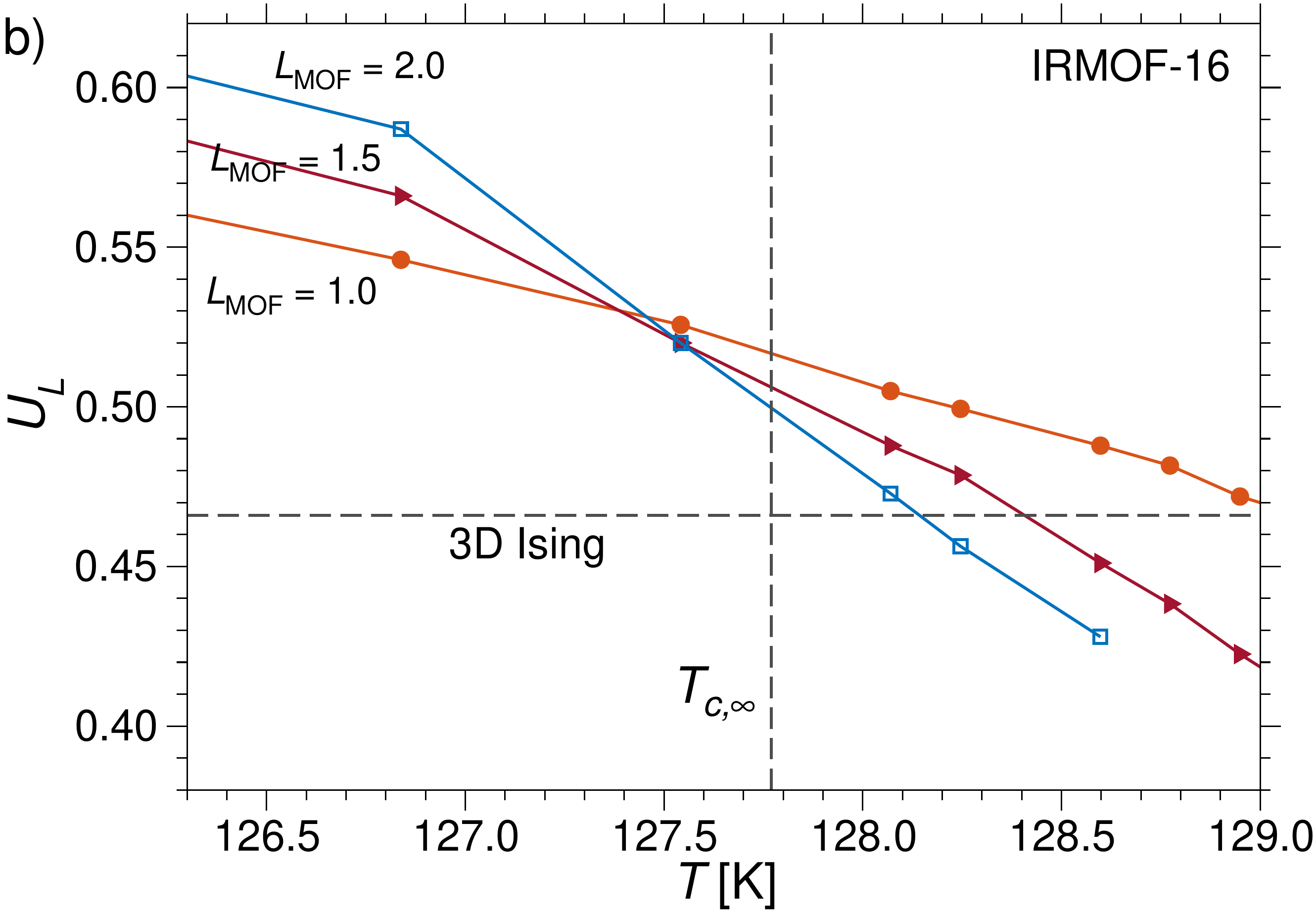}
\caption{\label{fig8} a) Binder's cumulant $U_L$ for the Ising-MOF system
for different system sizes $L$ close to the critical temperature. The
inset shows the intersection value of $U_L$, $U_L^\text{*}$, as a function
of system size and fits extrapolating $U_L^\text{*}$ to $L\to\infty$
using a linear function (red) and a constant (green). The figure in b)
shows the same cumulant for IRMOF-16. 3D Ising universal values of
$U_L^\text{*} \approx 0.465$ are included as indicated.}
\end{figure}
\begin{figure}
\includegraphics[width=0.48\textwidth]{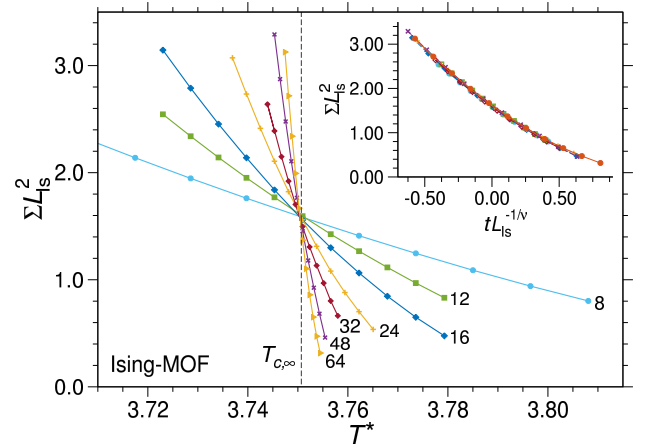}
\caption{\label{fig9} Surface tension, $\Sigma$ multiplied by $L_{\rm
Is}^2$ for all system sizes, as indicated by the numbers for $L_{\rm
Is}$. In the main panel $\Sigma L_{\rm Is}^2$ is shown as function of
temperature $T^*$ and in the inset as function of the scaling variable
$tL^{1/\nu}$ with $t=\frac{T-T_c}{T_c}$ using the 3D Ising universal
value $\nu=0.63$.}
\end{figure}
An appropriate order parameter ${\cal O}$ for the porous Ising model is
the difference between the magnetizations of the coexisting ferromagnetic
phases (${\cal O} = \Delta m$). A similar order parameter for the IRMOF-16
system is the density difference between the coexisting CH$_4$ fluids
(${\cal O} = \Delta \rho$). Approaching the critical point from below
along the binodal, the order parameter is expected to vanish as
\begin{equation}
 {\cal O} \propto \left(T_{c, L} - T \right)^{\beta},
 \label{eq_op}
\end{equation}
provided that the temperature difference $T_{c,L}-T$ is sufficiently
small.  Here, $T_{c, L}$ is the critical temperature corresponding
to a cubic system with linear dimension $L$. In the following, $T_{c,
L_{\rm Is}}$ and $T_{c, L_{\rm MOF}}$ denote the finite-size critical
temperatures for the Ising and the IRMOF-16 model, respectively. Using
the appropriate exponent $\beta$ in Eq.~(\ref{eq_op}), one would obtain
a straight line in a plot of ${\cal O}^{1/\beta}$ vs.~$T$ that intersects
the $x$-axis at $T_{c, L}$. The critical temperature in the thermodynamic
limit, $T_{c,\infty}$, can be determined via \cite{ferrenberg91}
\begin{equation}
 T_{c, L} = T_{c, \infty} + A L^{-1/\nu}, 
 \label{eq_tcl}
\end{equation}
with $A$ a critical amplitude and $\nu$ the critical exponent that
describes the divergence of the correlation length.  Figure \ref{fig8}
shows the rectification plots for the order parameter using the
value $\beta=0.326$ \cite{hasenbusch10}, predicted for the 3D-Ising
universality class.  As insets, the scaling plots for the critical
temperatures according to Eq.~(\ref{eq_tcl}) are displayed. Here, the
3D-Ising value $\nu = 0.63$ is used \cite{hasenbusch10}. The scaling works
nicely for the Ising model, resulting in the estimate $T_{c, \infty}
= 3.7507$.  The rectification plots for the IRMOF-16 system indicate
strong deviations from a straight line for small values of $(\Delta
\rho)^{1/\beta}$. This is very likely due to higher-order corrections
to the finite-size scaling prediction, Eq.~(\ref{eq_op}), see also
Ref.~\cite{ferrenberg91}. For the atomistic system, the estimate for
the critical temperature in the thermodynamic limit is $T_{c, \infty}
= 127.8$\,K.  The critical temperature can be also obtained from the
Binder cumulant $U_L$ \cite{binder81}. For the Ising model, $U_L$
is defined by Eq.~(\ref{eq_binder_ising}).  In Fig.~\ref{fig8}a,
the cumulant $U_L$ for the Ising model is plotted as a function of
temperature for different values of $L$. In the finite-size scaling
regime of the isotropic 3D bulk Ising model, $U_L$ for different $L$
is expected to intersect at the critical temperature and a universal
value $U_L^* =0.61069$ for the 2D Ising \cite{kamienarz93} and
$U_L^* =0.4655$ for the 3D Ising universality class \cite{luijten02}.
From Fig.~\ref{fig7}a small corrections to finite-size scaling can be
inferred. However, we can extrapolate $U_L^\text{*}$ to the thermodynamic
limit $L\to \infty$. Plotting $U_{L_i}^\text{*}$, as a function of
$L^{-\omega}$ and extrapolating this using a linear approximation via
$U_L^\text{*}(x=L^{-\omega}) = mx+c$ by fitting the parameters $m$ and
$c$. Such extrapolation can be found in the inset of Fig.~\ref{fig8}a,
also including an extrapolation using a constant, as it appears to work
equally well. In both cases we observe a deviation to $U_L^\text{*}$
as obtained by \cite{hasenbusch10} for the 3D Ising universality class,
which allows us to conclude the external framework potential introduces
(small) corrections to $U_L^\text{*}$ compared to the bulk system.
For the IRMOF-16 system, the Binder cumulant is defined by $U_L = 1 -
\frac{\langle\rho^{4}\rangle}{3\langle\rho^2\rangle^2}$, with $\langle
\rho^2 \rangle$ and $\langle \rho^4 \rangle$ being respectively the
second and fourth order moments of the probability distribution $P(\rho)$,
$\langle \rho^n \rangle = \int d\rho \rho^n P(\rho)$ with $n=2$ and $n=4$,
respectively.  As Fig.~\ref{fig7}b indicates, due to the small system
sizes the corrections to the finite-size scaling regime are much stronger
for the CH$_4$-IRMOF-16 system. Nevertheless, one can conclude also in
this case that the behavior of the cumulants is at least consistent
with 3D-Ising universality (however, the MOF geometry \emph{might}
change its value, as we discussed below Eq.~\eqref{eq_binder_ising}).

Finally, we present a more refined finite-size scaling
analysis for the Ising-MOF system, from which our most
accurate results follow (the so-called quotients method, see
Refs.~\cite{nightingale76,ballesteros96,amit05}).  Unfortunately, the
atomistic MOF systems that we can simulate are far too small to reproduce
this analysis.

The starting point is identifying a dimensionless scaling function. In
our case, the easiest to compute (and also the most accurate one) is the
free-energy cost of introducing a system-wide interface, namely $\Sigma
L^2$, see Eq.~\eqref{eq:Sigma-def}. Finite-size scaling tells us that
$\Sigma L^2$ scales as
\begin{equation}
\label{eq:nuFSS}
\Sigma L ^2 = g\Big(L^{1/\nu} (T-T_\mathrm{c})\Big)+{\cal O}(L^{-\omega}) \, ,
\end{equation}
where $g$ is a smooth scaling function and $\omega$ is the universal
leading corrections to scaling exponent.  Therefore, barring scaling
corrections, if we plot $\Sigma L^2$ as a function of $T$ for several
system sizes as we do in Fig.~\ref{fig9}, the curves will cross at
$T_\mathrm{c}$. Alternatively, if we represent $\Sigma L^2$ as a function
of $L^{1/\nu} (T-T_\mathrm{c})$ data from different system sizes should
collapse onto a master curve, see the inset of Fig.~\ref{fig9}.

In order to perform a precision computation of $T_\mathrm{c}$ and the
critical exponents, we consider pairs of lattices of sizes $(L_1,L_2)$. We
shall fix their ratio $s=L_2/L_1$ and consider the limit of large
$L=L_1$.  The corresponding curves $\Sigma L_1^2$ and $\Sigma L_2^2$,
see Fig.~\ref{fig9}, cross at a temperature scaling as
\begin{equation}
T^{(L,sL)}=T_\mathrm{c} 
+ A \frac{1-s^{-\omega}}{s^{1/\nu}-1}L^{-(\omega+\frac{1}{\nu})}\,,
\end{equation}
where $A$ is an amplitude and we have considered only the leading
corrections to scaling~\cite{binder81}. We have mostly considered $s=2$
for the ratio of system sizes.  This equation is used to obtain another
independent estimate of the critical temperature.

As for the critical exponents, let us consider a generic quantity
$O$ that, in the $L=\infty$ limit, diverges as $\langle O\rangle\sim
1/|T-T_\mathrm{c}|^{x_O}$. Finite size scaling implies that $\langle
O\rangle_L = L^{x_O/\nu} g_O(L^{1/\nu} (T-T_\mathrm{c})$, where $g_O$
is an unknown but smooth scaling function. Then it is easy to see that
the ratio evaluated at the crossing point $T^{(L,sL)}$ scales as
\begin{equation}
\label{eq:coc}
\frac{\langle O\rangle_{sL}(T^{(L,sL)})}{\langle
O\rangle_{L}(T^{(L,sL)})}=s^{x_O/\nu}+A_O L^{-\omega}\,.
\end{equation}
In the above expression, $A_O$ is an amplitude, and we have kept only
the leading corrections to scaling.

\begin{table}
\begin{tabular}{|c|c|ccc|}
\hline\hline
& bulk~\cite{hasenbusch10} &Model 1 & Model 2 & Model 3\\\hline
$\beta_\mathrm{sim}$& & 0.2666 & 0.2339 & 0.2564\\
\hline
$\beta_\mathrm{c}$& 0.22165463(8) &0.266642(7) & 0.233961(6) & 0.256359(5)\\
$h_\mathrm{c}$    & 0 &-0.563515(2)& -0.114187(5)& -0.0566572(7) \\
\hline
$\nu$    & 0.63002(10) & 0.629(9)  & 0.629(5) & 0.628(5) \\
$\eta$   & 0.03627(10) & 0.027(14) & 0.03(3)  & 0.04(7)  \\
$(\Sigma L^2)^*$ &$-$     & 1.57(3) & 1.58(8)  & 1.600(19)  \\
\hline
\hline
\end{tabular}
\caption{Extrapolation of the critical points and exponents to the
thermodynamical limit for the three models studied. These results are
obtained with tethered MC simulations containing at least $N_m=184$ points
at $\beta_\mathrm{sim}\sim \beta_\mathrm{c}$, and using the reweighting
method~\cite{ferrenberg88} to extrapolate to nearby values of $\beta$. The
exponents are extracted using the quotients method, see
Eq.~\eqref{eq:coc}. For the bulk universality class, we should also recall
the most precise known results $\nu=0.629971 (4)$ and $\eta=0.036298(2)$
that were obtained with the conformal bootstrap~\cite{kos16} (this technique
cannot provide $\beta_\mathrm{c}$).
\label{tab:criticalpointIsing}}
\label{tab:ana}
\end{table}
We employ Eq.~\eqref{eq:coc} with the derivative $\partial_T\Sigma$
($x_{\partial_T\Sigma}=1-2\nu$), and with $(\Delta m)^2$, see
Fig.~\ref{fig7}a ($x_{(\Delta m )^2}=2-\eta$ ($\eta$ is the anomalous
dimension, see e.g.~\cite{amit05}). The results of this analysis are
summarized in Table \ref{tab:criticalpointIsing}. In the table we present
results for the main model studied here and include as well exponents
for two modified versions of the Ising-MOF model, one with doubled
periodicity $P$ and with decreased spin-spin coupling constant $J$. As
expected, in both cases the phase behavior becomes more bulk-like with
respect to the critical temperature shift and the external field at the
critical point, $h_c$.

\subsection{Conclusions}
\label{sec4}
In this work, an Ising model for the adsorption of gas molecules in
metal-organic frameworks (MOFs) has been presented. The phase behavior
of this model has been directly compared to an atomistic simulation of
methane (CH$_4$) in IRMOF-16. The MOF-Ising model consists of frozen
lines of equal spins that are arranged such that they form an ordered
network with a cubic framework structure. Although the pores of the
proposed Ising model are extremely narrow (note that in our model 1 the
lines of frozen spins appear with a periodicity $P=4$), it exhibits a
line of first-order order transitions, each with the coexistence of {\it
bulk} phases, i.e.~ferromagnetic phases, that extend over the unit cells
of the framework structure.  A qualitatively similar phase behavior is
found for the atomistic MOF system and thus the MOF-Ising model can be
considered as a minimal model for the adsorption of gas molecules in MOFs.

The line of first-order transitions both in the MOF-Ising model and the
atomistic CH$_4$-IRMOF-16 system ends in a critical point. Consistent
with the observation of first-order transitions with coexisting
three-dimensional bulk phases, one may conjecture that the critical
behavior of the MOF systems belongs to the 3D Ising universality
class. However, this would imply that there is a divergent correlation
length that grows over the unit cells of the framework structure when
approaching the critical point. The existence of such a critical
behavior has hardly any counterpart in other systems with 3D Ising
universality. So it is a non-trivial issue whether the conjecture of
3D Ising behavior associated with the adsorption transitions in MOFs
holds. For the atomistic MOF system, the accessible system sizes are
too small to convincingly confirm the latter conjecture. However,
for the MOF-Ising system we have rationalized in this work that the
critical behavior is consistent with 3D Ising behavior. To this end,
we have performed a detailed finite-size scaling analysis of the order
parameter, the Binder cumulant and the surface tension.

The proposed MOF-Ising model can be easily extended to describe also
other phenomena, associated with the gas adsorption in MOFs. The IS
transition, where the coexisting phases form on the surface of the
framework structure, can be realized by an inhomogeneous distribution
of the interaction parameter $J$ describing the interaction between a
frozen and an active spin. Frozen up-spins sitting in the corners of the
framework structure (``metallic centers'') shall have stronger attractive
interactions with the active spins than the rest of the framework up-spins
on the ``linkers''. In this manner, one could stabilize a phase with an
enrichment of active up-spins at the corners of the framework and thus
there is the possibility of a phase transition from the latter phase to
one where there is an enrichment of active up-spins around the surface of
the whole framework. Another interesting theme would be the investigation
of phase-ordering kinetics in MOF-Ising models.  Due to the framework
of frozen spins, the domain coarsening in MOF systems is expected to be
very different from that in typical bulk fluids. The MOF-Ising model
is very well suited for the study of phase-ordering kinetics since it
allows to consider relatively large length and time scales and, as a
consequence, the scaling behavior and the morphology of the coarsening
dynamics could be investigated in detail.  All these issues shall be
addressed in forthcoming studies.

\begin{acknowledgments}
We thank Christoph Janiak for useful discussions. The authors
acknowledge financial support by Strategischer Forschungsfonds (SFF)
of the University of D\"usseldorf in the framework of the PoroSys
network and by the German DFG, FOR 1394 (grant HO
2231/7-2). V.M.M. and B.S. were partially supported by MINECO (Spain)
through Grant No.  FIS2015-65078-C2-1-P.  This project has received
funding from the European Union's Horizon 2020 research and innovation
program under the Marie Sk{\l}odowska-Curie grant agreement
No. 654971.  Computer time at the ZIM of the University of
D\"usseldorf is also gratefully acknowledged.
\end{acknowledgments}
%

%

%

%\bibliographystyle{apsrev4-1}
%\bibliography{shortjournal,references}
%
\end{document}